\documentclass[prd,aps,nofootinbib,preprintnumbers,floatfix,twocolumn]{revtex4}
\pdfoutput=1
\usepackage{amsmath,amssymb,url}
\newcommand{\ave}[1]{\left\langle #1 \right\rangle}

\usepackage{graphicx}

\begin{document}
\title{Near Zone Navier-Stokes Analysis of Heavy Quark Jet Quenching in
an $\mathcal{N}$ =4 SYM Plasma}

\author{Jorge Noronha$^{1,\,2,\,3}$, Giorgio Torrieri$^2$, and Miklos Gyulassy$^1$}
\affiliation{$^1$Department of
Physics, Columbia University, 538 West 120$^{\,th}$ Street, New York,
NY 10027, USA\\
$^2$Institut
f\"ur Theoretische Physik, J.W. Goethe--Universit\"at,
Max-von-Laue-Str. 1, 60438 Frankfurt am Main, Germany\\
$^3$Frankfurt
Institute for Advanced Studies, J.W. Goethe--Universit\"at,
Ruth-Moufang-Str. 1, 60438 Frankfurt am Main, Germany}

\begin{abstract}
The near zone energy-momentum tensor of a supersonic heavy quark jet
moving through a strongly-coupled $\mathcal{N}=4$ SYM plasma is
analyzed in terms of first-order Navier-Stokes hydrodynamics. It is
shown that the hydrodynamical description of the near quark region
worsens with increasing quark velocities. For realistic quark
velocities, $v=0.99$, the non-hydrodynamical region is located at a
narrow band surrounding the quark with a width of approximately
$3/\pi T$ in the direction parallel to the quark's motion and with a
length of roughly $10/\pi T$ in the perpendicular direction. Our
results can be interpreted as an indication of the presence of
coherent Yang-Mills fields where deviation from hydrodynamics is at
its maximum. In the region where hydrodynamics does provide a good
description of the system's dynamics, the flow velocity is so small
that all the nonlinear terms can be dropped. Our results, which are
compatible with the thermalization timescales extracted from
elliptic flow measurements, suggest that if AdS/CFT provides a good
description of the RHIC system, the bulk of the quenched jet energy
has more than enough time to locally thermalize and become encoded
in the collective flow. The resulting flow pattern close to the
quark, however, is shown to be considerably different than the
superposition of Mach cones and diffusion wakes observed at large
distances.

\end{abstract}


\date{\today}
\maketitle

\section{Introduction}

One of the most prominent experimental discoveries made at the
Relativistic Heavy Ion Collider (RHIC) has been the suppression of
highly energetic particles \cite{BRAHMS,PHENIX,PHOBOS,STAR}, which
suggests that the matter created at RHIC is a color-opaque, high
density medium of colored particles where fast partons quickly lose
energy by gluon emission \cite{lpm1,lpm2,brems2,cher1}. Measurements
of anisotropies in soft particle momentum distributions
\cite{BRAHMS,PHENIX,PHOBOS,STAR} have further indicated that soft degrees of freedom are approximately thermalized.  The degree of thermalization has been found to be considerably above the
predictions obtained within perturbative quantum chromodynamics
(QCD) \cite{gyulvisc} and, in fact, it seems to be compatible with
the ``perfect fluid'' scenario where the strongly-coupled
quark-gluon plasma (sQGP) has almost zero viscosity
\cite{gyulassymclerran,heinz,shuryak,teaney,heinz2,romatschke}. This indicates that perturbative methods may fail at describing the
microscopic properties of the matter created in ultrarelativistic
heavy ion collisions.

An alternative approach for understanding jet suppression in the
sQGP relies on analyzing the similar problem of a heavy quark moving
through an $\mathcal{N}=4$ super Yang-Mills (SYM) plasma. There the
Anti-de Sitter/Conformal Field Theory (AdS/CFT) correspondence
\cite{maldacena} provides a way to compute observables
at strong coupling on the CFT side, such as the stress tensor of the
$\mathcal{N}=4$ SYM  plasma, starting from gravitational calculations
in five-dimensional $\rm{AdS}$ space.

It is important to remark that one has to be careful when comparing QCD and $\mathcal{N}=4$
SYM given that both the symmetries and the microscopic degrees of
freedom are very different. The hope is that some of the differences become
unimportant in the strongly coupled limit where the system is to a
good approximation a locally thermalized liquid.

A notable result obtained using the AdS/CFT correspondence
\cite{etatos1} was that the ratio between the viscosity $\eta$ and
entropy density $s$ of an $\mathcal{N}=4$ SYM
plasma in the limit of large number of colors $N_{c}$ and large 't Hooft coupling $\lambda$ converges to a ``universal limit''
\begin{equation}
\frac{\eta}{s} = \frac{1}{4\pi}\, ,
\end{equation}
which has been conjectured \cite{etatos2} to be the lowest value for
this ratio obtainable in a physical system. Remarkably, hydrodynamic
models find that the viscosity of the system created at RHIC is
compatible, or even lower, than this limit \cite{heinz,heinz2,romatschke}.

The apparent local thermalization at RHIC suggests the possibility
that the energy lost in jets will itself become locally thermalized,
i.e., encoded in the local hydrodynamic flow. In the immediate vicinity of the jet the variation of the thermodynamic quantities is probably very large, which means that a hydrodynamic description of the system is bound to fail in this region. The typical length scale associated with this region is, for instance, given by the sound attenuation length $\Gamma=4\eta/3\omega < 1/\pi T$, where $\omega$ is the enthalpy of the medium (see \cite{CasalderreySolana:2006sq}). For distances scales larger than $\Gamma$, one expects that the system can be described in terms of non-linear viscous hydrodynamics. Sufficiently far from the jet, however, the flow velocity and the gradients of the thermodynamic quantities are so small that linearized hydrodynamic equations should provide an accurate description of the system's dynamics. In the linearized limit, for certain energy deposition patterns, this gives rise to Mach cones, which can be readily connected to the fluid's
equation of state \cite{lifshitzlandau}. In the absence of these
conditions, the flow pattern is likely to be considerably more
complex but will nevertheless result in a correlation between the
flow and the jet. The possibility of detecting these effects in
heavy ion collisions has recently been subject to intense
theoretical and experimental interest
\cite{Mach1,Mach2,Mach3,Mach4,Machexp1,Machexp2,Machexp3}.

The degree of thermalization of the energy of the suppressed
jets is a fundamental probe of the microscopic properties of the
matter created at RHIC. Because of the apparent strong coupling of
the system, non-perturbative methods such as AdS/CFT seem to be
necessary. Recently, this method has been used to obtain the energy-momentum tensor of an
infinitely heavy quark moving at constant velocity through a strongly
coupled plasma \cite{gubser1,gubser2}. The results have been
analyzed in the far field limit \cite{Gubser:2007ga,gubser2,gubser3,gubsermeson,yaffe1,yaffe2}
in terms of linearized Navier-Stokes hydrodynamics and, encouragingly, it has
been shown that structures resembling Mach cones and diffusion
wakes, with an angle comparable to that expected from the equation
of state, do form in the far-field limit. In fact, in \cite{yaffe2} it was shown that linearized Navier-Stokes hydrodynamics provides a very accurate description of the system at distances $\geq 5/\pi T$ (in the limit of the large $N_c$ and $\lambda$).

As mentioned above, close to the quark linearized hydrodynamics, and eventually the
hydrodynamic description, should fail to account for the
energy-momentum tensor of the system. Distinguishing between
non-linear, turbulent hydrodynamics, and the failure of
hydrodynamics is impossible within the linearized regime. In this work we focus on the region very close to the quark and perform an analysis of the {\em near quark } energy-momentum tensor computed in \cite{yarom} (see also \cite{Gubser:2007nd} and \cite{Gubser:2007xz}) using viscous,
non-linear (first-order) Navier-Stokes hydrodynamics. We construct the flow vector
by boosting the energy-momentum tensor in \cite{yarom} to the local
rest frame of the flow. This flow vector is used to construct the
energy-momentum tensor corresponding to a viscous fluid having an
equation of state and transport coefficients identical to those of
strongly coupled $\mathcal{N}=4$ SYM matter with the same flow
vector. We then quantitatively examine the discrepancies between the two tensors and plot them as a
function of the distance from the heavy quark.

Throughout this paper the units are $\hbar=c=k_{B} = 1$ and 4-vectors are denoted by capital
letters, e.g., $U^{\mu} = (U^{0},\vec{U})$. Also, the Minkowski
metric $g_{\mu\nu}={\rm diag}(-,+,+,+)$ is used.

\section{The near-quark energy-momentum tensor}

The energy-momentum tensor of a heavy quark passing through an
$\mathcal{N}=4$ SYM plasma at finite temperature $T$
can be computed by considering metric fluctuations due to a string that
is hanging down from the boundary of an AdS Schwarzschild (AdS-SS) background
geometry \cite{gubser2} \footnote{Fluctuations of the string can be neglected when $M/T\gg \sqrt{\lambda}$, where M is the quark's mass \cite{Herzog:2006gh,Mach3,CasalderreySolana:2007qw}. In this paper we consider only this classical limit. Assuming that $T=0.5$ GeV and $\lambda=3\pi$, one sees that the classical string dynamics is a much better approximation for bottom quarks than for charm quarks.}. In fact, in this limit the total action that describes the supergravity approximation to type IIB string theory in an AdS-SS background and the classical string is given by the sum of the following partial actions
\begin{equation}
\ A_G = \frac{1}{16\pi G_5 }\int \sqrt{-G}\left(R+\frac{12}{L^2} \right)
\label{gravityaction}
\end{equation}
and
\begin{equation}
\ A_{NG} = -\frac{1}{2\pi \alpha' }\int \sqrt{-G_{\mu\nu}^{(0)}\partial_{\alpha}X^{\mu}\partial_{\beta}X^{\nu}}d^2\sigma,
\label{nambugoto}
\end{equation}
where $L$ is the radius of $AdS_5$, $G_5=\pi L^2 /2N_{c}^2$, $\alpha'=L^2 /\sqrt{\lambda}$, $G_{\mu\nu}$ is the total metric, and $G_{\mu\nu}^{(0)}$ is the metric of the unperturbed AdS-SS black hole (without effects from the string), which can be obtained from
\begin{equation}
\ ds^2 = \frac{L^2}{z^2}\left(-g(z)dt^2+d\vec{x}^{\,2}+\frac{dz^2}{g(z)}\right)
\label{lineelement}
\end{equation}
where $z$ goes from $0$ at the AdS boundary to $z_0 =1/\pi T$ at the black hole horizon ($T$ is the Hawking temperature associated with the black hole) and $g(z)=1-(z/z_0 )^4$. The string coordinates $X^{\mu}(\sigma,\tau)$ in the Nambu-Goto action in Eq.\ (\ref{nambugoto}) are chosen in such a way that the string endpoint (which corresponds to the heavy quark in the 4-dimensional boundary) moves at constant speed $v$ and no energy flows from the horizon into the string \cite{Herzog:2006gh,gubser4}.

Minimizing the action $S$ with respect to $G$ leads to the full set of Einstein's equations. It is sufficient for our purposes here to consider instead the linearized Einstein's equations for the metric fluctuations $h_{\mu\nu}$, which are defined via $G_{\mu\nu}=G_{\mu\nu}^{(0)}+h_{\mu\nu}$. It can be shown \cite{gubser2,yarom} that the contribution from the moving quark to the total energy-momentum tensor is $T_{quark}=\frac{1}{\pi}\sqrt{\frac{\lambda}{1-v^2}}Q$, where the tensor $Q$ is obtained by expanding $h$ in powers of $z$ near the boundary, i.e., $h\sim Q\,z^4$. The total energy-momentum tensor in the lab frame that describes the near-quark dynamics of the plasma was computed by Yarom \cite{yarom} and it reads
\begin{equation}
\ T_{\mu\nu}^{Y}=P_0\,{\rm diag}\{3,1,1,r^2\}+\Delta T_{\mu\nu}(x_{1},r).
\label{energymomentumtensor1}
\end{equation}
where the explicit form of $\Delta T_{\mu \nu}$ is
\begin{equation}
 \Delta T_{tt}  =  \alpha \frac{v  \left(r^2(-5+13v^2-8v^4)+(-5+11v^2)x_1^2\right)x_1}{72\left(r^2(1-v^2)+x_1^2\right)^{5/2}}, \nonumber \\
\end{equation}
\begin{equation}
 \Delta T_{t x_1}  =  -\alpha \frac{v^2 \left(2 x_1^2+(1-v^2)r^2\right)x_1}{24 \left(r^2(1-v^2)+x_1^2\right)^{5/2}}, \nonumber \\	
\end{equation}
\begin{equation}
 \Delta T_{tr}  =  -\alpha \frac{(1-v^2) v^2 \left(11 x_1^2+8r^2(1-v^2)\right) r}{72 \left(r^2(1-v^2)+x_1^2\right)^{5/2}}, \nonumber \\	
\end{equation}
\begin{equation}
 \Delta T_{x_1 x_1}  = \alpha \frac{v  \left(r^2(8-13 v^2 +5v^4)+(11-5v^2)x_1^2\right)x_1}{72 \left(r^2(1-v^2)+x_1^2\right)^{5/2}}, \nonumber \\
\end{equation}
\begin{equation}
 \Delta T_{x_1 r}  =  \alpha \frac{ v (1-v^2) \left(8 r^2(1-v^2)+11 x_1^2\right)r}{72 \left(r^2(1-v^2)+x_1^2\right)^{5/2}}, \nonumber \\
\end{equation}
\begin{equation}
 \Delta T_{r r}  = -\alpha \frac{v (1-v^2)  \left(5 r^2(1-v^2) + 8 x_1^2\right)x_1}{72 \left(r^2(1-v^2)+x_1^2\right)^{5/2}}, \nonumber \\
\end{equation}
\begin{equation}
 \Delta T_{\theta \theta}  = -\alpha \frac{r^2\,v (1-v^2) x_1 }{9 \left(r^2(1-v^2)+x_1^2\right)^{3/2}}. \label{yarmom7}
\end{equation}
and $\alpha=\gamma_q\sqrt{\lambda}\,\pi^2 T^4$. In the equations above we used the dimensionless cylindrical coordinates $x_1 =X_{1}\,\pi T$ (where $X_1 =X-vt$ is the comoving coordinate of the quark), $r=X_{p}\,\pi T$, and also defined $\gamma_{q}=1/\sqrt{1-v^2}$. Moreover, $P_0=\left(N_{c}^2-1\right) \pi^2 T^4/8+\mathcal{O}(N_{c}^0)$ is the pressure of the ideal SYM plasma \cite{Gubser:1996de}. Note that the disturbances in the medium caused by the string are parametrically of $\mathcal{O}(T^2/X^2)$ in the near-quark region \cite{yarom}, while the vacuum Coulomb stress (not included here) is $\mathcal{O}(1/X^4)$ \cite{gubser2}.

It is assumed throughout the derivation of Eq.\ (\ref{energymomentumtensor1}) that the metric disturbances caused by the moving string are small in comparison to the $AdS_{5}$ background metric. Therefore, this result is correct as long as this condition is fulfilled. In fact, since $\Delta T_{\mu \nu}$ scales inversely with the total distance from the quark, the region where the condition $\Delta T_{\mu \nu}/P_0 < 1$ (or, equivalently, $h$ small in comparison to $G^{(0)}$) holds can be taken to be arbitrarily small as long as the limit where $N_c\to \infty$ and $\lambda$ is large is employed. However, in order to evaluate the relevance of this approach to heavy ion collisions, one may choose $N_{c}=3$, $\lambda=3\pi$ ($\alpha_s =0.25$), and $\gamma_q =10$. Using these parameters, one obtains that the magnitude of the energy density disturbances caused by the string along the jet axis, $\Delta T_{tt}/\left(3P_0\right)|_{r=0}$, is $<1$ when $|X_1| > 1/(\pi T)$ (which is comparable to uncertainty principle bound on the mean free path discussed in \cite{gyulvisc}).

\section{Comparison to Hydrodynamics}

We shall now compare the tensor in Eq. \ref{energymomentumtensor1} to the one that represents a solution to the relativistic Navier-Stokes equations with an arbitrary flow.  We underline that the {\em quantitative} values for the flow and thermodynamic quantities resulting from our analysis should be taken with caution. In fact, one would expect that the validity of a low-energy hydrodynamic description of the near-field region worsens as one becomes closer to the heavy quark. However, this analysis is necessary to ascertain the domain of validity of (general) hydrodynamics, as opposed to non-equilibrium field configurations (up to the ``opposite extreme'' of fully anisotropic classical solutions to the field equations).  As we show in this paper and subsequent work \cite{redneck}, ascertaining this domain of validity of hydrodynamics is crucial for the development of the phenomenological properties of AdS/CFT based models.

In hydrodynamics, the Knudsen number $K_N$ is defined as the ratio between the mean free path $l_{MFP}$ and a characteristic spatial dimension of the system $q$. Hydrodynamics is applicable when $K_N \equiv l_{MFP}/q \ll 1$. In conformal field theories at finite temperature, the only dimensionful parameter is given by the temperature $T$ and, thus, both $l_{MFP}$ and $q$ should be proportional to $1/T$. However, the mean free path is not a well defined quantity in $\mathcal{N}=4$ SYM theories at very strong coupling. Nevertheless, one can still define an effective Knudsen number as
\begin{equation}
\ K_N \equiv \Gamma\,\Big |\frac{\vec{\nabla}\cdot \vec{S}}{S}\Big |
\label{knudsen}
\end{equation}
where $\Gamma=1/3\pi T$ is the sound attenuation length (to leading order in $N_c$), $S_i=-T_{0i}^{Y}$ is the momentum density, and $S=\sqrt{\vec{S}^2}$. Note that this quantity does not depend on $N_c$ and $\lambda$ but it strongly depends on $v$. It is illustrated in Fig.\ \ref{knudsen} how $K_N$ changes with $v$. For $v=0.75$, $K_N \ll 1$ practically everywhere but in a small region of radius $\sim 1/\pi T$ that surrounds the quark. However, when $v$ takes more realistic values, say, $v=0.99$, $K_N \gtrsim 1$ in a narrow region located at $-3/\pi T < X_1 <3/\pi T$ and $X_p >0$. Therefore, one can expect that in this region the system cannot be described by hydrodynamics.

\begin{figure}[t]
\includegraphics[width=8cm]{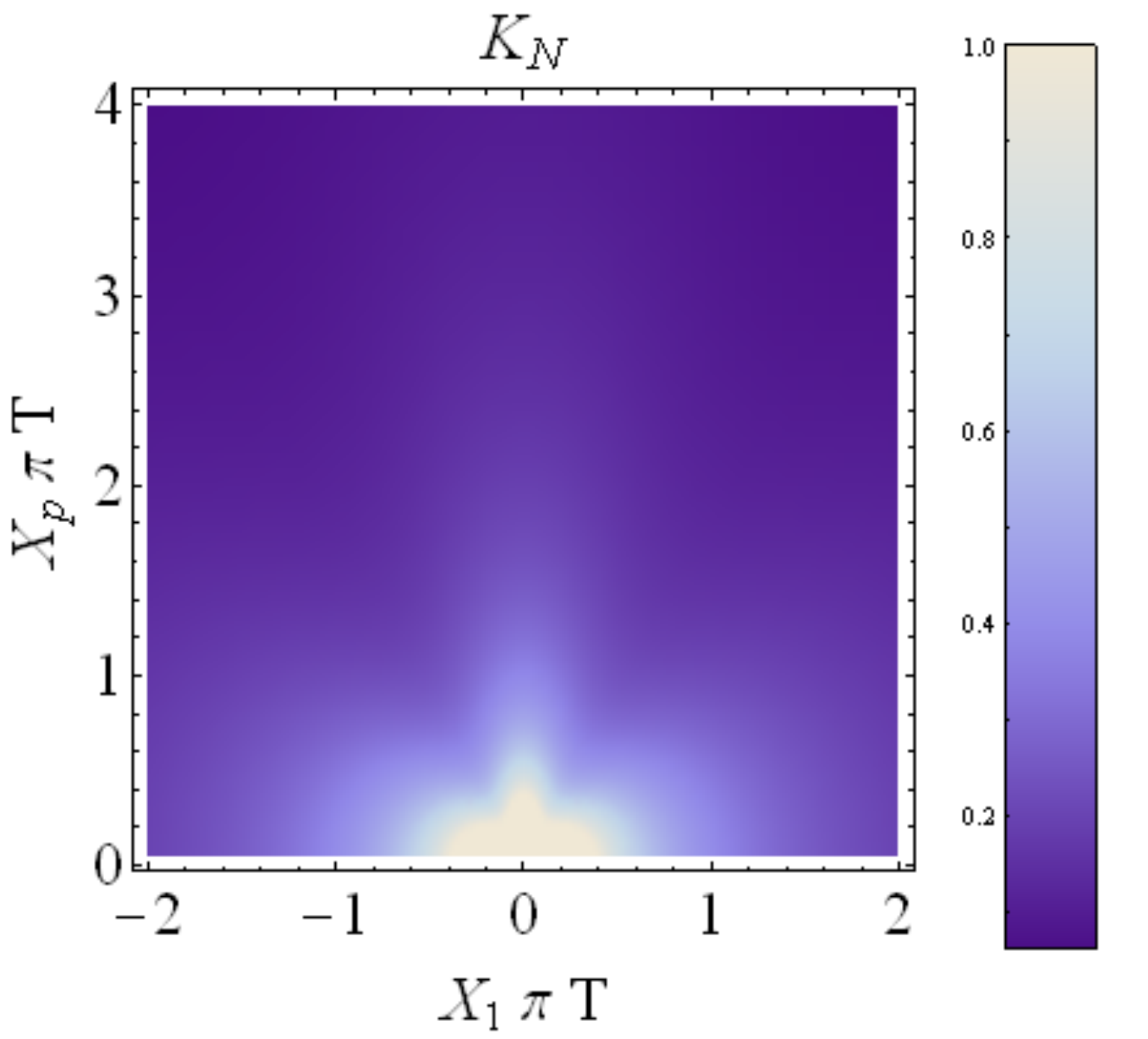}\\~\\~\\~\\
\includegraphics[width=8cm]{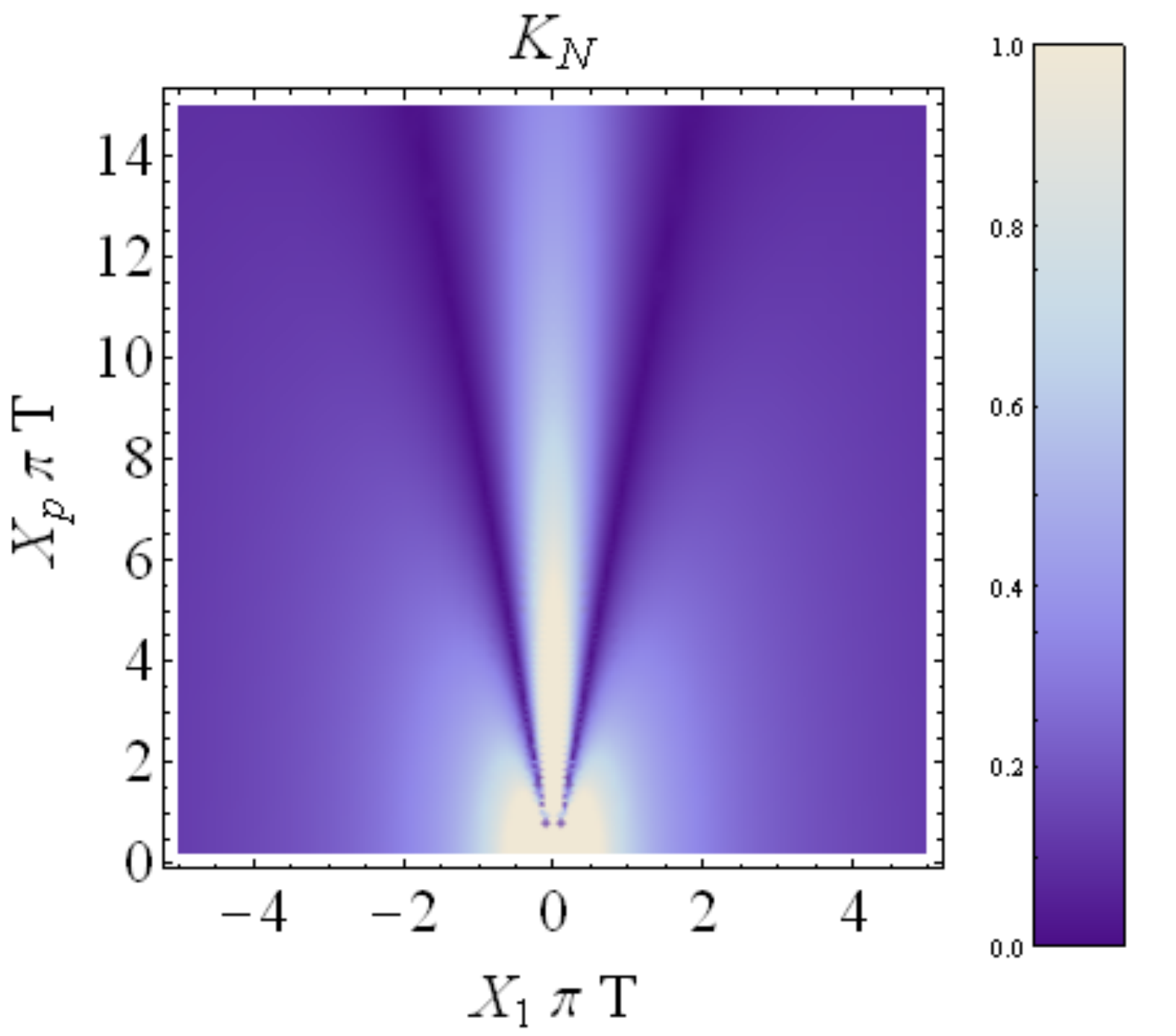}
\caption{\label{knudsen} (Color online) Effective knudsen number for $v=0.75$ (left panel) and $v=0.99$ (right panel). Note that different scales were used in the left and right panels. Ratio values outside of the limits were color-coded as the limits.}
\end{figure}

Another way to verify whether the system described by the energy-momentum tensor in Eq.\ (\ref{energymomentumtensor1}) admits a hydrodynamic description is to compute its flow vector $U^{\mu}$, which can be obtained by boosting to the Landau frame where $(T_{0 i}^{Y})_{L} = 0$ (denoted henceforward by brackets and subscript $L$, $\left(...\right)_L$). Note that, unless the system is a {\em coherent} field where the phase velocity is equal to the speed of light (such would be the case of an electromagnetic wave), this transformation is always possible. It can be accomplished by solving a system of two equations for the two space-like components of $U^\mu$, $U_{1}$ and $U_{r}$ ($U_{\theta}=0$),
\begin{equation}
\left( T_{0 i}^{Y} \right)_L = \Lambda^{\mu}_{i}\,T_{\mu \nu}^{Y}\,\Lambda^{\nu}_{0} = 0 ,
\label{landauequation}
\end{equation}
where $\Lambda^{\mu}_{i}$ is a general coordinate dependent Lorentz transformation
\begin{equation}
\Lambda = \left( \begin{array}{cc} \gamma &  \vec{U}^{T} \\
\vec{U} & \,\,\,\,1+\frac{\vec{U}\otimes\vec{U}^{T}}{\vec{U}^{2}}\left(
\gamma -1 \right)
\end{array}
 \right),
\end{equation}
where $\gamma\equiv U^0=\sqrt{1+\vec{U}^2}$. Using the representation
\begin{equation}
\ T^{Y} = \left( \begin{array}{cc} \varepsilon &  -\vec{S}^{T} \\
-\vec{S} & \,\,\,\,\hat{\tau}
\end{array}
 \right)
\end{equation}
where $\varepsilon=T_{00}^{Y}$ and $\hat{\tau}_{ij}=T_{ij}^{Y}$ ($i,j=1,2,3$), we obtain that Eq. (\ref{landauequation}) becomes
\begin{equation}
\vec{U}=\frac{1}{\left(\gamma\varepsilon-\vec{S}^T \cdot \vec{U}\right)}\left[1+  \frac{\vec{U}\otimes\vec{U}^{T}}{\vec{U}^{2}}\left(
\gamma -1 \right)  \right] \left(\gamma \vec{S}-\hat{\tau}\vec{U}\right).
\label{numericalsolutionflow}
\end{equation}
For finite $N_c$ and $\lambda$ this equation can only be solved numerically. However, for very large $N_c$ and large $\lambda$ one can use that, to leading order in $1/N_c$, $\gamma\approx 1$, $\varepsilon\approx 3P_0$, and $\hat{\tau}\approx P_0$. Using these approximations, one can then obtain that
\begin{equation}
\vec{U}\approx\frac{\vec{S}}{4P_0}.
\label{largeNcflow}
\end{equation}
In deriving the equation above we used that $\sqrt{\lambda}/N_{c}^2\ll 1$. The fact that $|\vec{U}| \sim \mathcal{O}(\sqrt{\lambda}/N_{c}^2)$ in the large $N_c$, $\lambda$ limit implies that nonlinear terms in the hydrodynamic description are subleading contributions that can be neglected. Thus, if the system's dynamics can be described by hydrodynamics, these equations have to be linear for the present string theory setup obtained in the large $N_c$, $\lambda$ limit \cite{chesler}. However, at finite $N_c$ and $\lambda$ nonlinear effects are expected to be relevant. These nonlinear effects can only be properly taken into account by incorporating subleading $1/N_c$ corrections. Because these corrections have yet to be calculated, we shall extrapolate the validity of our leading order results and make a comparison between the near-quark energy-momentum tensor in Eq.\ (\ref{energymomentumtensor1}) and a hydrodynamic ansatz using the phenomenologically relevant set of parameters $N_c=3$ and $\lambda=3\pi$. In this work, to estimate both the non-linearities and non-equilibrium effects in the near-zone, we shall use the non-linear ansatz in Eq. (\ref{numericalsolutionflow}) rather than Eq. (\ref{largeNcflow}).

The energy-momentum tensor of a viscous fluid containing no
conserved charges can be parametrized using the energy density
$\rho$, pressure $p$, and energy flow vector $U^{\mu}$
\cite{lifshitzlandau}
\begin{equation}
\label{tmunu} T^{NS}_{\mu \nu} = (\rho + p )\,U_\mu U_\nu +
p\,g_{\mu \nu} + \Pi_{\mu \nu}
\end{equation}
(in our metric convention $U_{\mu} U^\mu=-1$ ). $\Pi^{\mu \nu}$ is the shear tensor, which contains the
corrections to local thermalization that can be expanded in terms
of the gradients of $U^\mu$. To first-order in these gradients
(Navier-Stokes limit), this tensor is given by \cite{lifshitzlandau}
\begin{eqnarray}\label{shear}
\Pi^{\mu \nu} &=& -\eta\,(\partial^\mu U^\nu +
\partial^\nu U^\mu + U^\mu U_\alpha \partial^\alpha U^\nu \nonumber \\
&+&  U^\nu U^\alpha \partial_\alpha
U^\mu)+\frac{2}{3}\eta\,\Delta^{\mu\nu}\left(\partial_\alpha
U^\alpha\right),
\end{eqnarray}
where $\Delta^{\mu\nu}=g^{\mu\nu}+U^{\mu}U^{\nu}$ is the local
spatial projector and $\eta=\left(N_{c}^2-1\right) \pi\, T^3 /8$. Note that we kept
the nonlinear terms in $U^{\mu}$ in the definition of
$\Pi_{\mu\nu}$.

Once the solution to $\vec{U}$ has been found (by solving the
transcendental equation (\ref{numericalsolutionflow}) numerically), the shear tensor $\Pi_{\mu
\nu}$ and the energy-momentum tensor $T^{NS}_{\mu \nu}$ can be
computed from Eqs.\ (\ref{shear}) and (\ref{tmunu}).  The
continuity of $\Pi_{\mu \nu}$ confirms the stability of our
numerical solution and the existence of only one physically relevant
set of $\vec{U}$.

For an inviscid fluid, the form of the transformed tensor will be
\begin{equation}
\left( T_{\mu \nu}^{id} \right)_L = {\rm diag}\{\rho,p,p,p\}.
\label{idtensor}
\end{equation}
If the first-order Navier-Stokes approximation provides a full description of the
system,
\begin{equation}
\left( T_{\mu \nu}^{Y} \right)_L = \left( T_{\mu \nu}^{NS} \right)_L = \left( T_{\mu \nu}^{id} \right)_L + \left( \Pi^{\mu \nu} \right)_{L}
\end{equation}
where
\begin{itemize}
\item $\left( \Pi^{\mu \nu} \right)_{L}$ is the shear tensor of Eq.\ (\ref{shear}) transformed to the Landau frame.
\item $\rho$ in Eq.\ (\ref{idtensor}) is given by
\begin{equation}
\rho(x_1, r)= \left(T^Y_{00} \right)_L - \left( \Pi_{00} \right)_L
\label{Too0}
\end{equation}
\item The equation of state is the CFT one, $p=\rho/3$.
\end{itemize}
Note that Eq.\ (\ref{idtensor}) can in general be very different than the
``background'' tensor $\ave{T_{plasma}}=P_0\,{\rm diag}\{3,1,1,r^2\}$, since it
is defined in the frame {\em comoving} with the flow generated by
the jet.

The deviation from Navier-Stokes hydrodynamics can then be
quantitatively investigated by defining the tensor
\begin{equation}
Z_{\mu \nu} =  T_{\mu \nu}^{Y}  -\Pi_{\mu\nu}  \label{zdef}
\end{equation}
and studying, in particular, discrepancies between $\left(Z^{\mu \nu}\right)_L$ and $\left(
T_{\mu \nu}^{id} \right)_L$. If $\left( T_{\mu \nu}^Y \right)_L$ corresponds to a solution of the
Navier-Stokes equations, we should get
\begin{equation}
\left(Z_{11}\right)_L = \left(Z_{22}\right)_L = \left(Z_{33}\right)_L = \frac{1}{3} \left(Z_{00}\right)_L,
\end{equation}
and
\begin{equation}
\left(Z_{ij}\right)_L=0.
\end{equation}

\section{Results}

In this preliminary analysis, we will concentrate on the case where
$\lambda=3\pi$, $N_{c}$=3, and the quark velocity in the lab frame
is $v=0.99$. We would like to remark at this point that $1/\pi T$ is
the natural uncertainty principle bound on spatial resolution at
temperature $T$ in an ultrarelativistic plasma and provides a lower
bound on the mean free path and hence viscosity \cite{gyulvisc}.
Within a radius $1/\pi T$ of the quark jet our semiclassical
dissipative fluid analysis is expected to fail and non-equilibrium
and field coherence effects may dominate the dynamics.

\begin{figure}[t]
\begin{centering}
\includegraphics[width=8cm]{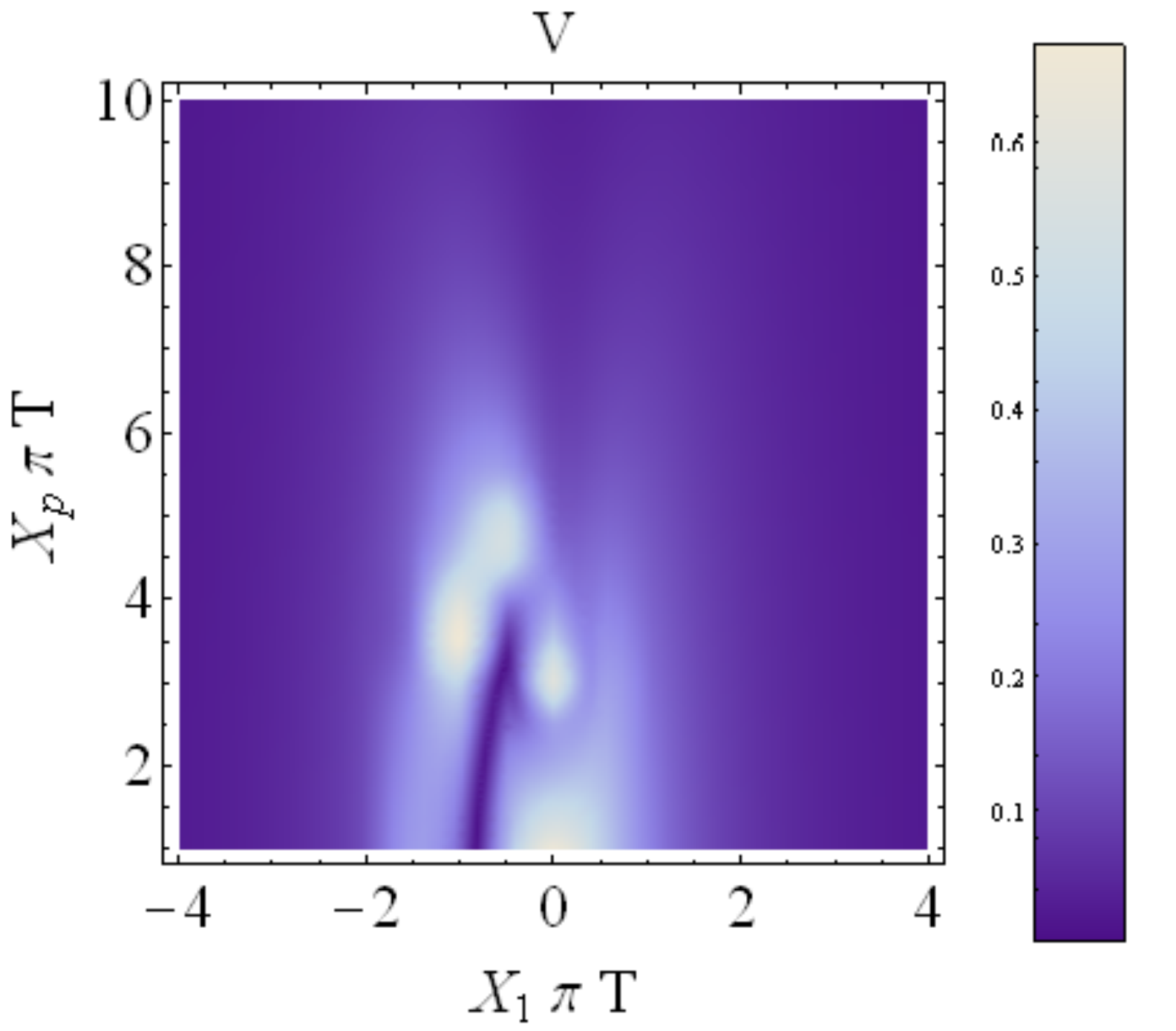} \caption{\label{fig1}
(Color online) Contour plot of the magnitude of the non-relativistic flow velocity $\vec{V}$ for $v=0.99$.}
\end{centering}
\end{figure}

The plot seen in Fig. \ref{fig1} immediately illustrates that the flow nearby the quark is very different than a superposition of a Mach cone and a diffusion wake ($\vec{V}$ is such that $\vec{U}=\gamma\vec{V}$). Note that $\vec{V}^2 < 1$ everywhere. This structure is expected to be somewhat stable with respect to $v$ since viscosity results in a length scale below which the fluid ``sticks'' to and co-moves with the quark. In a steady state solution, even a small viscosity can produce an observable disturbance {\em preceding} the quark. In the hydrodynamic simulations of \cite{betzpol,betzjets,betzjets2} the fluid is ideal. Hence, the ``shock''-like nature of the quark disturbance is recovered and the energy density is qualitatively different.

\begin{figure}[t]
\begin{centering}
\includegraphics[width=8cm]{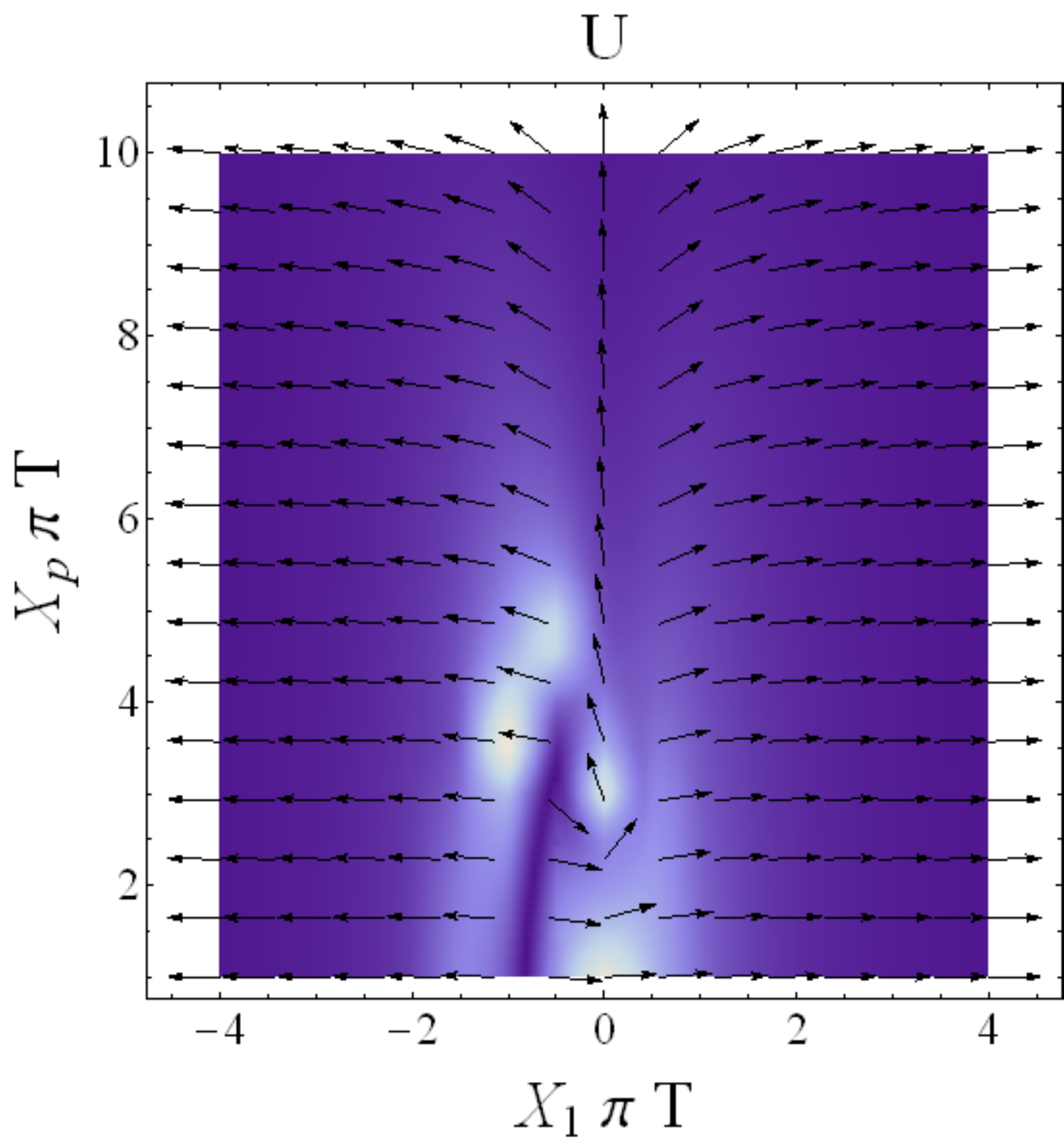} \caption{\label{fig2}
(Color online) Contour plot of the collective flow $\vec{U}$ for $v=0.99$.}
\end{centering}
\end{figure}

In Fig. \ref{fig2} we superimposed the magnitude and the vector plot of the collective flow $\vec{U}$.  As the
quark moves forward, matter is being transported by collective flow
in the direction opposite to quark motion. This effect is, in
itself, not completely surprising since perturbative calculations
\cite{wangprivate} admit both momentum flow out of the quark (if
repulsive channels dominate) and into the quark (if attractive
channels dominate). Furthermore, in a system with no conserved
charges, collective flow and heat diffusion can compensate
each other since the effect of both is moving energy-momentum
throughout the system. This picture, is however very far from the
point-like energy source considered in the ``text-book'' Mach cone
example of \cite{lifshitzlandau}.

\begin{figure}[t]
\begin{centering}
\includegraphics[width=8cm]{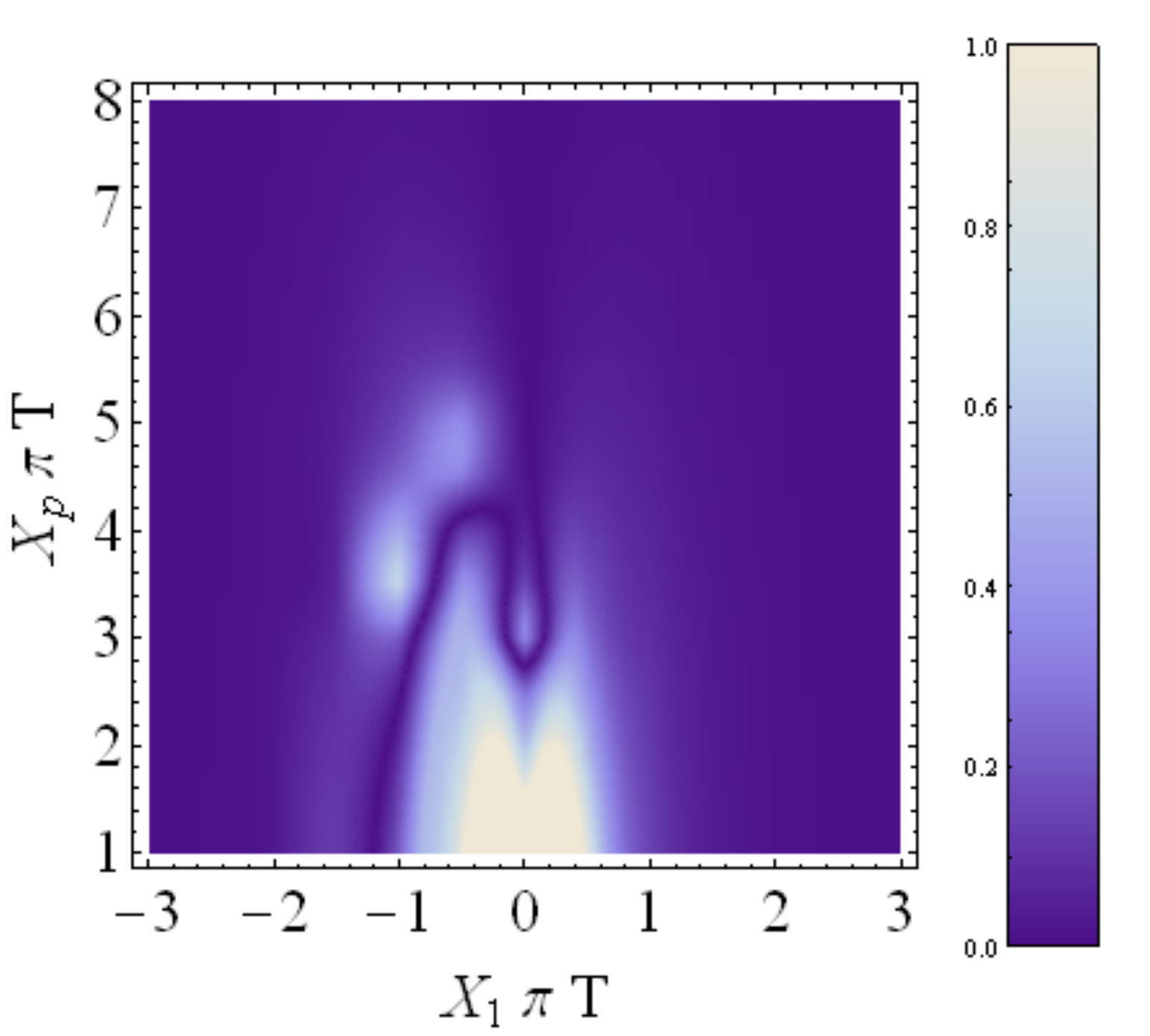}
\caption{\label{fig3} (Color online) Comparison between the full
numerical solution for $\vec{U}$ and the leading order expression in
Eq.\ (\ref{largeNcflow}), i.e., $|\vec{U}-\vec{S}/4P_0 |$ for $v=0.99$.}
\end{centering}
\end{figure}

In Fig.\ \ref{fig3} we compare the full numerical solution of Eq.\
(\ref{landauequation}) and the leading order result in Eq.\
(\ref{largeNcflow}). This plot indicates that the system's dynamics
becomes linear outside the narrow region between  $-1/\pi
T<X_1<1/\pi T$ and $0<X_p <6/\pi T$. This is in agreement with the
result in Fig.\ \ref{fig1} because the magnitude of $\vec{V}$ (or
$\vec{U}$) becomes very small outside this region.

\begin{figure*}
\includegraphics[width=8cm,clip=]{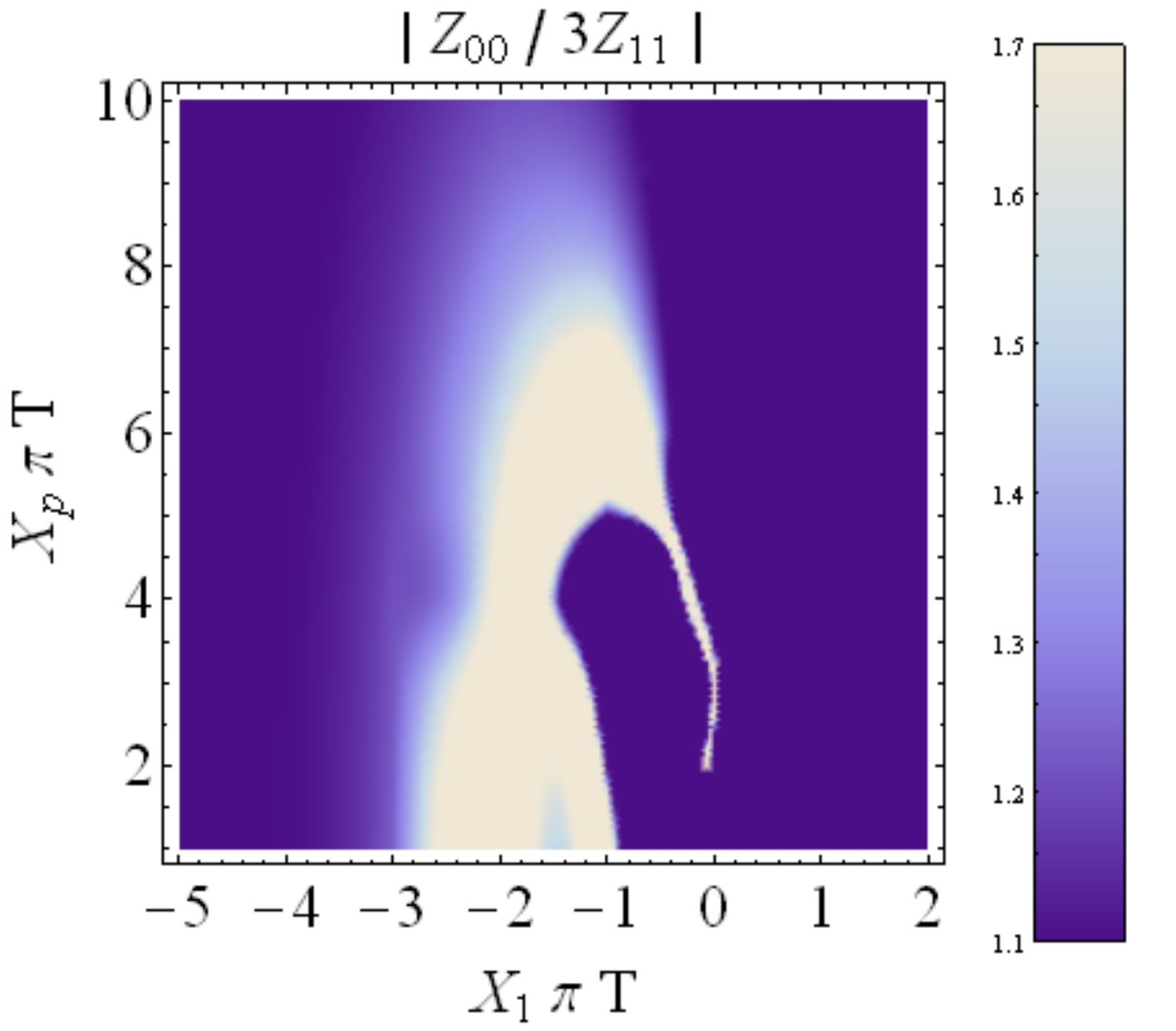}
\includegraphics[width=8cm,clip=]{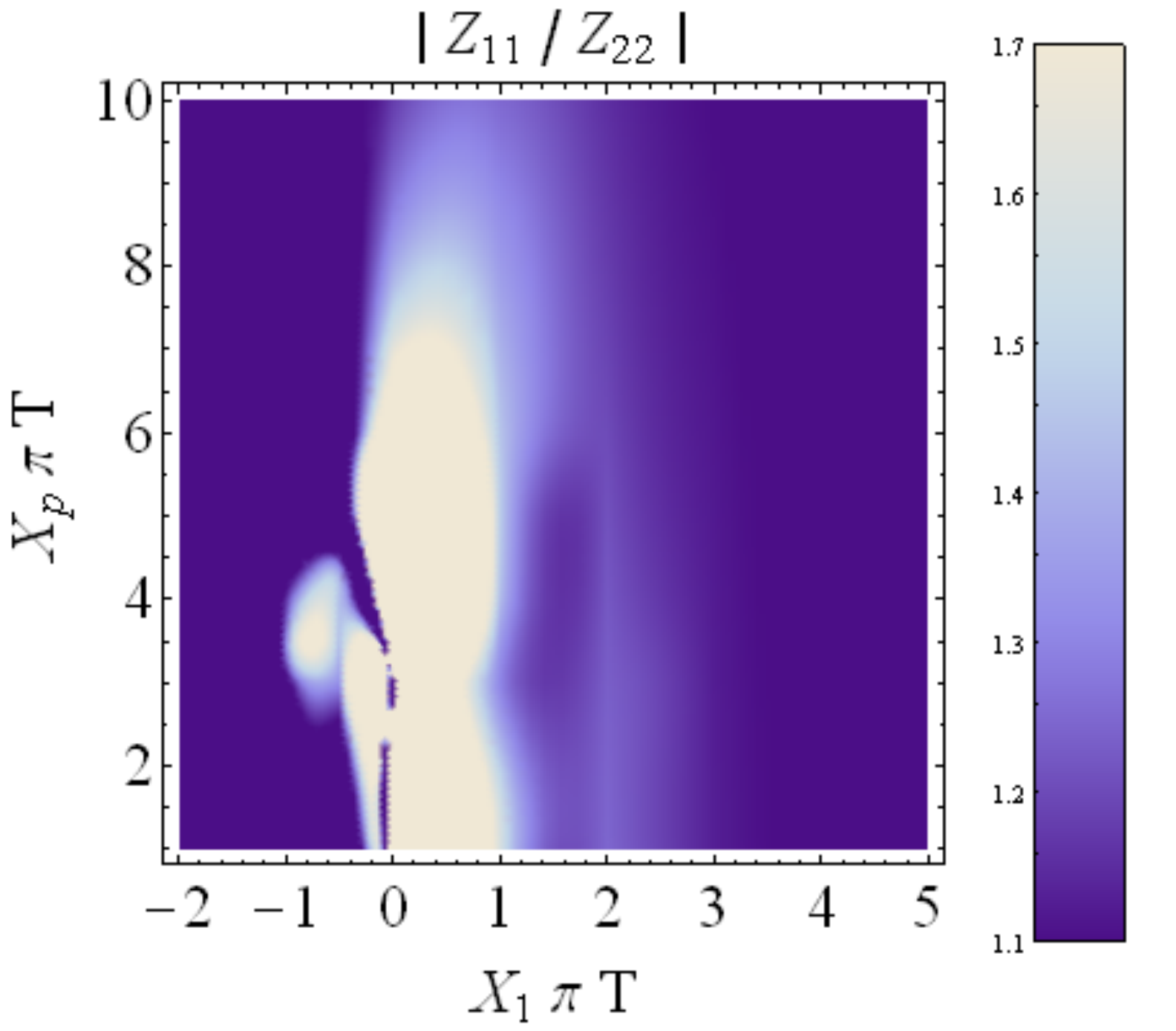}\\~\\~\\~\\
\includegraphics[width=8cm,clip=]{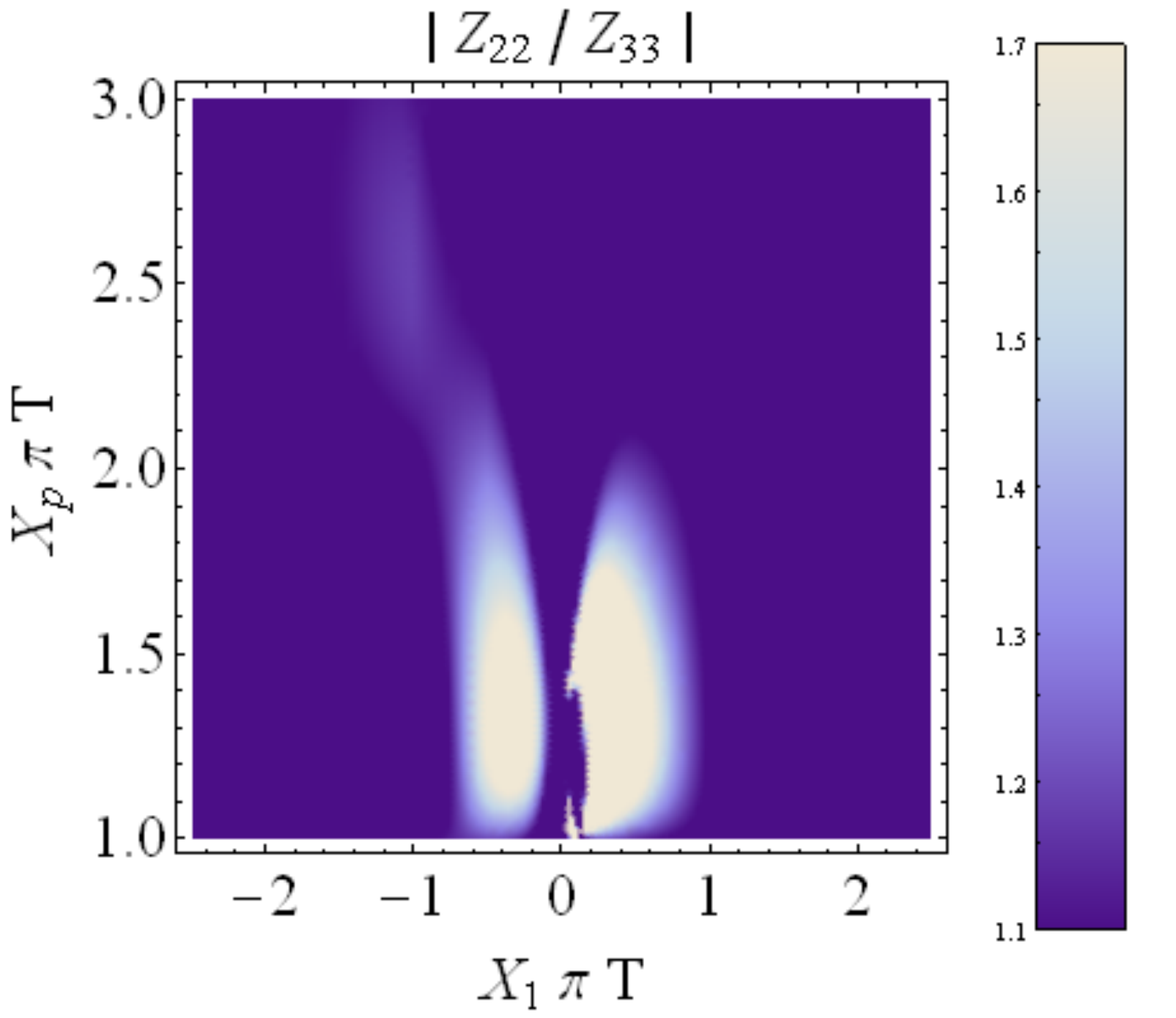}
\includegraphics[width=8cm,clip=]{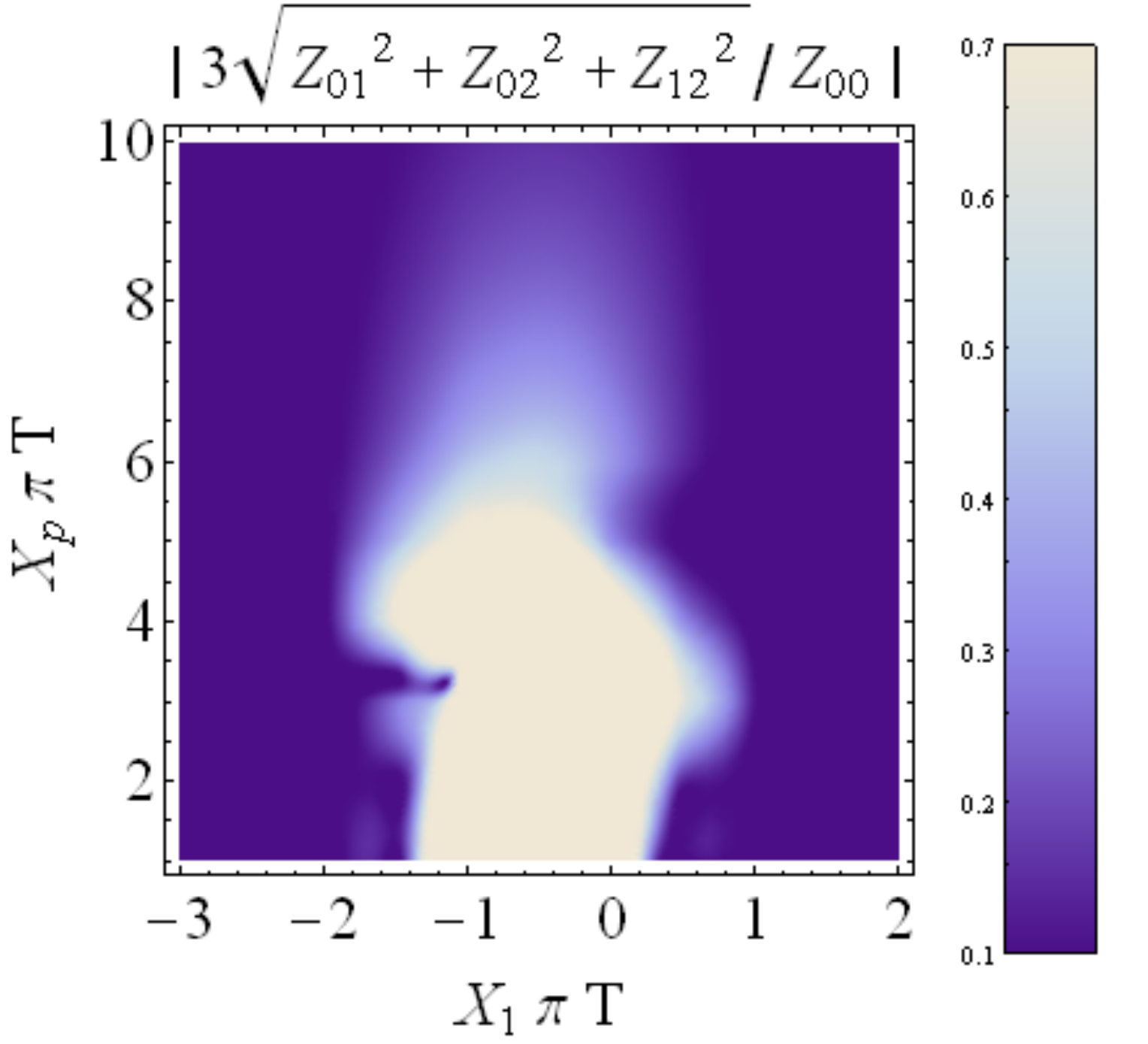}
\caption{\label{fig4} (Color online) Tests of local equilibrium for
the elements of the energy-momentum tensor in the Landau frame.
$Z_{\mu \nu}$ is defined in Eq. \ref{zdef} and $v=0.99$. Ratio values outside of
the limits were color-coded as the limits.}
\end{figure*}

Finally, Fig.\ \ref{fig4} shows the discrepancy between the
Navier-Stokes tensor and $T_{\mu\nu}^{Y}$ in the Landau frame, which
are parametrized by the tensor $Z_{\mu \nu}$ (Eq. \ref{zdef}). As
can be seen, the magnitude and structure of the discrepancy remains
somewhat the same across all components of $Z_{\mu \nu}$. Outside
the region where $-3/\pi T < X_1 < 3/\pi T$ and $0<X_p<10/\pi T$
(the quark is at $X_1=X_p=0$), the system can be described by the
first-order Navier-Stokes energy-momentum tensor to an accuracy of
roughly $90 \%$. Inside that region, however, the discrepancy
quickly rises and diverges. Note that in the region where
Navier-Stokes provides a good description of the system the
numerically calculated velocity flow $\vec{U}$ matches the leading
order in Eq.\ (\ref{largeNcflow}). This implies that, in this
region, the system is better described by the linearized version of
the first-order Navier-Stokes equations.

The detailed structure of the divergence is numerically difficult to
examine but its sudden onset suggests the presence of coherent
Yang-Mills fields, where deviation from hydrodynamics is at its
maximum. Coherent fields do not of course have a comoving frame, but
they can be transformed into a frame (which will in general be
different than the frame set by $U^{\mu}$ as calculated here) where
the energy-momentum tensor given by
\begin{equation}
\left( T_{\mu \nu}^{coherent} \right) = {\rm diag}\{\rho',p',0,0\}
\label{idtensorcoherent}
\end{equation}
(where $\rho'$ and $p'$ are taken to mean energy and momentum flow
density, not "true" thermal densities and pressures). Here, the
$X_{1}$ direction is taken to be parallel to the Poynting vector.
The $Z_{ii}$ calculated for this field can go to zero, which results
in divergences in ratios of $Z$'s as seen in Fig.\ \ref{fig4}. This
indicates that coherent Yang-Mills fields are present in the
vicinity of the quark. In fact, note that Yarom's $\Delta T_{\mu\nu}$ (\ref{yarmom7}) does seem to have a Lorentz contracted Weizs\"acker-William-like form but its $\mathcal{O}(T^2/X^2)$ parametric dependence is quite distinct from the vacuum Coulomb behavior \cite{redneck}.

\begin{figure}[t]
\begin{centering}
\includegraphics[width=8cm]{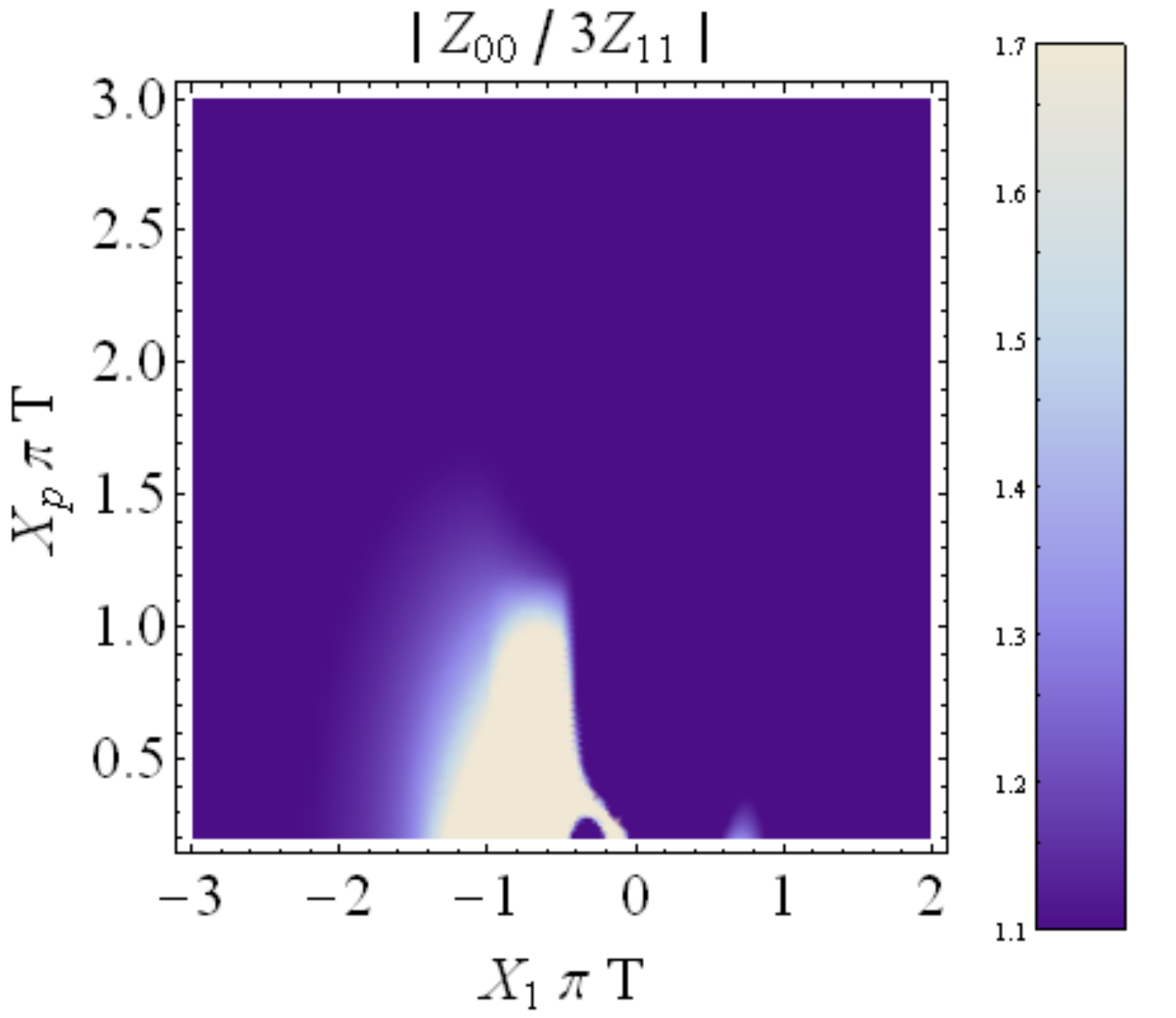} \caption{\label{fig5}
(Color online) Test of local equilibrium for the elements of the energy-momentum tensor in the Landau frame for $v=0.75$ and $\lambda=3\pi$. $Z_{00}$ is defined in Eq. \ref{zdef}. Ratio values outside of the limits were color-coded as the limits.}
\end{centering}
\end{figure}

Our results for the components of $Z$ do not significantly change if
we increase the intensity of the coupling to $\lambda=6\pi$.
However, at lower quark velocities the Navier-Stokes equations
provide an accurate description of the system's dynamics at
distances $\geq 1.5 /\pi T$ (see Fig.\ \ref{fig5}). These results
coincide remarkably well with the limit estimated through the
uncertainty principle \cite{gyulvisc}.

\begin{figure}[t]
\begin{centering}
\includegraphics[width=8cm]{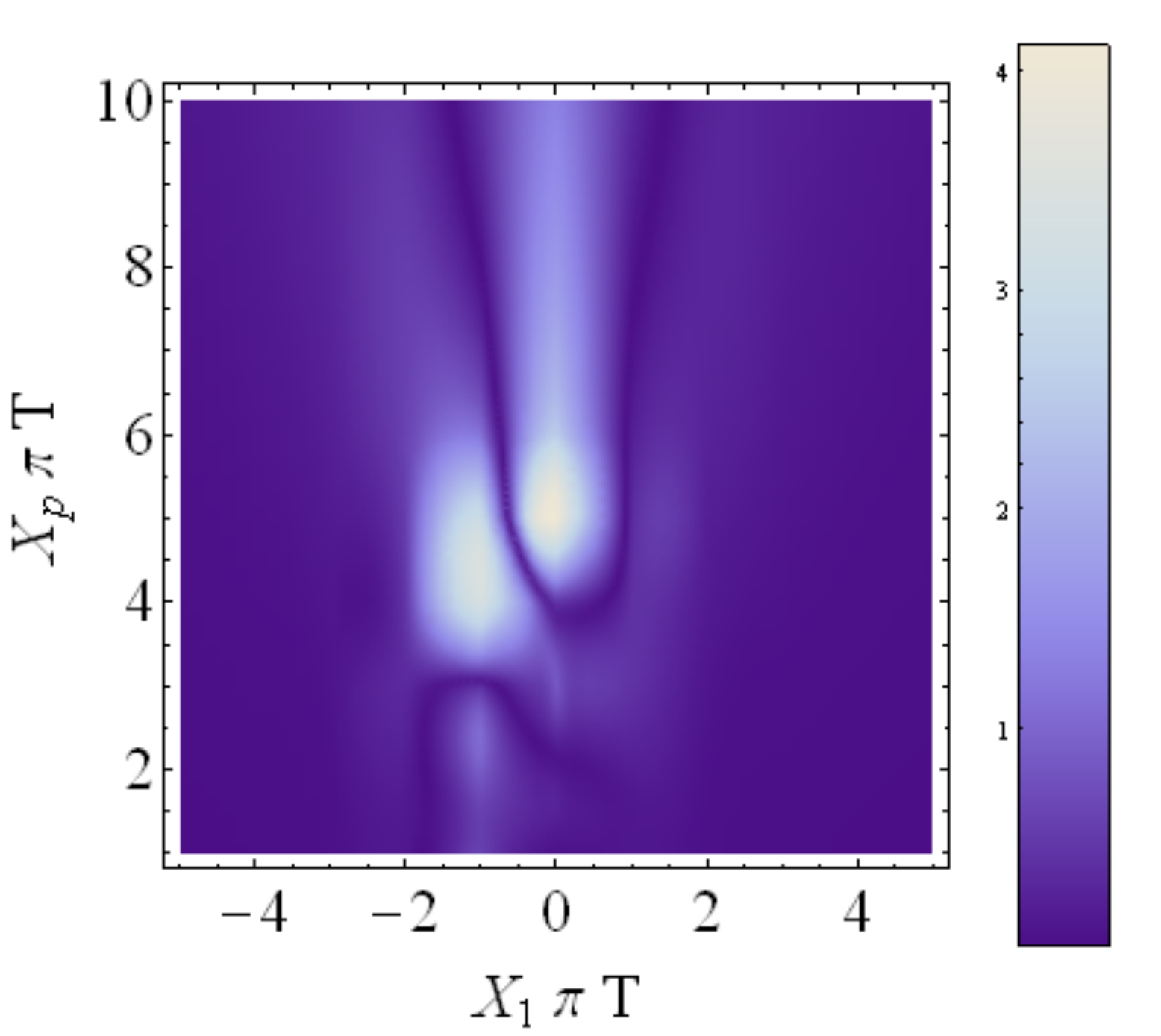} \caption{\label{fig7}
(Color online) The relative importance of the viscous tensor with respect to the energy-momentum tensor in the Landau frame, $\sqrt{\sum_{\mu\nu}\Pi_{\mu \nu }^2/\sum_{\mu\nu} Z_{\mu\nu}^2}$, corrected for double-counting. Here $v=0.99$.}
\end{centering}
\end{figure}

Fig.\ \ref{fig7} confirms that $\Pi_{\mu \nu}$ stays smooth and finite throughout, and the divergence is due to a maximal breakdown of isotropy (one of the $Z_{ii}$ going to zero). Furthermore, Fig.\ \ref{fig7} shows that in the regime in which hydrodynamics works, $\Pi_{\mu \nu}$ is a perturbatively small correction to the energy-momentum tensor. The regime where $\Pi_{\mu \nu}$ becomes significant approximately coincides with the regime in which the Navier-Stokes description stops to be a good description of the physical system. Thus, it is unlikely that higher gradient corrections to Navier-Stokes equations (such as \cite{israelstewart,koide,teaney2}) will make a significant difference in our results.

The contours of our figures look non-trivial and no easy interpretation of them can be given here, especially as they are highly susceptible to $1/N_c$ corrections (as remarked in the previous section). The fine detailed structure in the figures is sensitive to the interpolation numerics and only the global features are numerically stable. We remark, however, that the size of the ``inner region'' is very similar in all the figures and also agrees well with the estimate for the ``Knudsen boundary'' in Fig. \ref{knudsen}.  It is therefore worth underlining how the results agree with the expectations described in the previous sections. The system is neatly defined by a Knudsen region where $-3/\pi T < X_1 <3/\pi T$.  Inside this layer, non-equilibrium deviations from the hydrodynamic description dominate.  On the other hand, outside this region nearly ideal and linearized hydrodynamics provides an acceptable description of the system.

\section{Conclusions}

Encouragingly, the scale at which the energy deposited by the jet
thermalizes is comparable to the thermalization time constraints
obtainable from anisotropic flow measurements \cite{heinz}, which
suggests that in the strongly coupled regime the nearly inviscid
microscopic dynamics responsible for creating large $v_2$ such as
those seen at RHIC also has the potential to transform the energy of
quenched hard jets into flow of soft particles. The short timescale
seen in the present work indicates that, provided strongly coupled
AdS/CFT is indeed qualitatively similar to sQGP, the energy
deposited by jets in the system becomes to a very good approximation
locally thermalized.

Second-order hydrodynamic corrections \cite{israelstewart} may
somewhat change the results presented here. However, as argued
previously, the contribution of the relaxation time to collective
dynamics might be not so large. Analyzes of other systems within the
AdS/CFT framework, such as \cite{janik}, confirm this. An analysis
of the presently investigated solution within the second order
framework, analogously to that presented in \cite{janik,kovchegov}
for a Boost-invariant fluid, is currently in progress.

It should be noted that there is to date no universally agreed
framework for relativistic hydrodynamics beyond leading order, with
several higher order frameworks possible (see, for example,
\cite{koide,teaney2,israelstewart} and references within these
works). This still unresolved ambiguity might hold the key to
explaining some puzzles encountered in trying to model heavy ion
collisions using non-ideal hydrodynamics (the numerical
instabilities encountered in \cite{heinz2}, and the very low
viscosity encountered in \cite{romatschke}). For a discussion of
second-order viscous hydrodynamics in conformal field theories at
finite temperature see \cite{Baier:2007ix, Bhattacharyya:2007jc}.

We hope that AdS/CFT techniques might help in clarifying the
situation by providing a bonafide ``analytical'' solution for a
strongly coupled relativistic quantum system with none of the
approximations, whose validity is questionable at strong coupling,
that usually go into the derivation of transport equations. For such
a clarification, an analysis of AdS/CFT inspired solutions using
methods similar to those described here is necessary. This effort
could, independently from AdS/CFT's applicability to heavy ion
collisions, lead to a better comprehension of the general and
fundamental problem of the dynamics of strongly-coupled relativistic systems.

In conclusion, in this work we have compared the analytically
obtained energy-momentum tensor of a very massive quark moving through a
strongly coupled $\mathcal{N}=4$ SYM plasma with a non-ideal tensor
given by the first-order Navier-Stokes ansatz. We have found that
the energy deposited by the quark is already thermalized in the
region where $|X_1 | > 3/\pi T$ and $X_p >1/\pi T$, which defines a
length scale that is considerably shorter than the time-scale of the
jet traveling throughout the medium, or the medium's hydrodynamic
evolution. Thus, if strongly coupled $\mathcal{N}=4$ SYM is an
appropriate description of RHIC physics, we have every reason to
expect that in the strongly coupled region the missing jet energy
has been thermalized. The thermalization timescale is (indeed, the
lifetime of the heavy ion system), slightly smaller than the
timescale found in \cite{yaffe2}. Thus, in view of the results
present here, one can say that linearized first-order Navier-Stokes
hydrodynamics provides an accurate description of the
$\mathcal{N}=4$ SYM plasma surrounding the heavy quark down to
distances of about $3/\pi T$. A quantitative hydrodynamic analysis
of the energy deposited in jets at smaller distances needs to be
treated nonlinearly, as in \cite{betzjets,betzjets2}.

We expect that further progress in examining strongly coupled
$\mathcal{N}=4$ SYM systems will better clarify the thermalization
timescale and dynamics of strongly coupled field theories.

\section{Acknowledgements}

The authors thank A. Yarom for useful correspondence and D. Rischke
for useful suggestions. Comments from S. Gubser, J. Casadelrrey-Solana, and P. M. Chesler on the earlier version of this paper are gratefully appreciated. J.N. acknowledges support by the Frankfurt
International Graduate School for Science (FIGSS) and Gesellschaft f\"ur Schwerionenforschung (GSI). G.T. would like
to thank J.W.-Goethe Universit\"at and the Alexander Von Humboldt
foundation for the support provided for this research. M.G. and J.N. acknowledge partial support from DOE under Grant No. DE-FG02-93ER40764. M.G. also thanks DFG, ITP, FIAS for partial support.

\end{document}